\documentclass[a4paper]{article}
\pdfoutput=1

\usepackage{INTERSPEECH2021}

\title{TB or not TB? Acoustic cough analysis for tuberculosis classification}
\name{Geoffrey Frost$^1$, Grant Theron$^2$, Thomas Niesler$^1$}
\address{
  $^1$Department of Electrical and Electronic Engineering, Stellenbosch University, South Africa \\
  $^2$SAMRC Centre for Tuberculosis Research, DSI/NRF Centre of Excellence for Biomedical Tuberculosis Research, Stellenbosch University, South Africa
  }
\email{\{gfrost, gtheron, trn\}@sun.ac.za}

\begin{document}

\maketitle

\begin{abstract}
\vspace*{-1mm}
In this work, we explore recurrent neural network architectures for tuberculosis (TB) cough classification. In contrast to previous unsuccessful attempts to implement deep architectures in this domain, we show that a basic bidirectional long short-term memory network (BiLSTM) can achieve improved performance. In addition, we show that by performing greedy feature selection in conjunction with a  newly-proposed attention-based architecture that learns patient invariant features, substantially better generalisation can be achieved compared to a baseline and other considered architectures. Furthermore, this attention mechanism allows an inspection of the temporal regions of the audio signal considered to be important for classification to be performed. Finally, we develop a neural style transfer technique to infer idealised inputs which can subsequently be analysed. We find distinct differences between the idealised power spectra of TB and non-TB coughs, which provide clues about the origin of the features in the audio signal.
\end{abstract}
\vspace*{1mm}

\section{Introduction}
\vspace*{-1mm}

In 2021, 10 million people were reported to have developed tuberculosis (TB), of whom 1.5 million died. As a result, TB was the second most lethal infectious disease globally, ranking above HIV/AIDS and just below COVID-19 \cite{world2021global}. The majority of TB cases occur in developing nations where access to public health care is limited by complex socio-economic factors, making it difficult to identify and control the spread of the disease and resulting in patients receiving improper care \cite{foster2015economic}. 

Whilst published research covering cough classification is currently limited, a few studies have shown promising results when distinguishing between: wet and dry coughs \cite{amrulloh2016novel, swarnkar2013automatic}, pneumonia \cite{amrulloh2015cough, abeyratne2013cough}, and more recently COVID-19 \cite{pahar2021covid, mouawad2021robust, bagad2020cough}. Because TB is predominately a respiratory disease, it results in patients developing a chronic cough. It has been shown in previous work that it is possible to distinguish between the coughs of TB patients and healthy controls by utilising simple statistical classifiers \cite{botha2018detection}. 
More recently, these methods have been evaluated on a dataset that aims to reflect \textit{real-world} conditions, whereby coughers all suffer from some lung ailment that is in some cases TB~\cite{pahar2021automatic}. 
Whilst frequency bands important for classification were identified~\cite{botha2018detection}, a thorough investigation into the acoustic patterns being learnt has not yet been conducted. 
Moreover, work considering TB cough classification has relied on linear models utilising fixed dimensional inputs which are typically frame-wise averages of acoustic features. 
Thus, temporal information present in a cough has so far been disregarded.

In this work we show that recurrent deep learning architectures can be used successfully for TB cough classification, and improve upon existing methods. In addition, by incorporating an attention mechanism and a new loss term, combined with frugal feature selection, we show that model generalisation can be improved. Using the same attention mechanism, we are able to visualise the temporal regions of the feature space that are learnt to be important for cough classification. By considering idealised TB negative and TB positive coughs produced by a technique normally used for neural style transfer, we discuss the distinct characteristics of a TB cough captured by the neural network.

\section{Data}
\vspace*{-1mm}

We report classification results on a dataset comprising 74 individual patients and 1564 coughs. 
Previous work in TB cough classification has relied on relatively small datasets gathered in a single recording environment from a small number of patients. 
This is problematic when training deep-architectures due to their tendency to overfit, for example, to confounding socio-environmental factors which are especially important to disregard in a clinical setting \cite{kelly2019key}. 
Relying on recordings from a single environment restricts data diversity and consequently the final model's ability to generalise. In an attempt to address this, we combine the datasets used previously in~\cite{botha2018detection} and in~\cite{pahar2021automatic}, referred to as the Brooklyn and Wallacedene datasets respectively. 
This is in an effort to yield a more environmentally diverse dataset. 
Brooklyn was collected in a noise-isolated facility from patients known to have TB and healthy controls, whilst Wallacendene was collected in a noisy environment, from patients who all suffer from either TB or some other lung ailment (confirmed later by sputum analysis). 
This combined dataset is summarised in Table \ref{tab:dataset}.

\begin{table}[h!]
        \centering
        \caption{Dataset used for experimentation. TB and $\overline{\text{TB}}$ indicate TB positive and negative respectively.}
        \vspace*{-1mm}
         \begin{tabular}{l c c c}
          \toprule
           & TB & $\overline{\text{TB}}$ & Total\\
          \midrule
           Patients  & 28 & 46 & 74\\
           Total coughs & 844 & 720 & 1564\\
           Mean cough length (s) & 0.60 & 0.64 & 0.62 \\
           Std dev cough length (s) & 0.34 & 0.29 & 0.32\\
          \bottomrule
        \end{tabular}
     \label{tab:dataset}
\end{table}
\vspace*{-1mm}

\subsection{Cross-validation and testing}
\vspace*{-1mm}

We divide the combined dataset into a training set (which is further subdivided for cross-validation) and a test set, containing 49 and 25 patients respectively. Importantly, both the Brooklyn and Wallacedene datasets are represented equally in all splits. Furthermore, splits are performed patient wise, ensuring that all coughs originating from the same patient are only present in one set, and we ensure a uniform distribution of TB positive and negative (TB and $\overline{\text{TB}}$) patients across splits. The training set is further divided into 4 folds (each consisting of its own train and development set) for cross-validation using the same previously described procedure. 

\section{Models}
\vspace*{-1mm}

We first train and evaluate several binary classifiers, including: logistic regression (baseline), a basic BiLSTM and a BiLSTM with attention. Next, we use the attention-based architecture to deduce important cough characteristics by generating idealised coughs for each class through a neural style-transfer technique. A sequential forward search (SFS) \cite{ferri1994comparative} is performed for both recurrent architectures to identify the most important frequencies for classification and to investigate its impact on model generalisation. This information is considered in conjunction with the temporal regions identified as important for classification by analysis of the attention weights.

\subsection{Logistic regression}
\vspace*{-1mm}

Previous work in TB cough classification has focused on simple linear models since it was observed that complex neural networks resulted in degraded performance. In both \cite{botha2018detection} and \cite{pahar2021automatic}, logistic regression (LR) outperformed all other considered classifiers. As such, we use it as a baseline with which our architectures will be compared. LR is a simple approach that linearly models a probability $f(\mathbf{x})$ given a set of $d$ predictors $\mathbf{x}\in \mathbb{R}^{d}$ using  learnable parameters $\boldsymbol{\theta}$. This  highlights an important limitation of LR: each predictor $\mathbf{x}$ is a feature vector computed from a frame of audio. To obtain the probability that a cough is associated with TB, the average of the frame probabilities is computed. In doing so, any temporal information is lost.

\subsection{BiLSTM}
\vspace*{-1mm}

RNNs have successfully been used in several acoustic classification tasks. Acoustic feature vectors are processed sequentially, each updating the network's internal hidden states which contain complex context-rich information and are available at the next time step. This allows the network to learn temporal relations important to the task at hand.
In this work, we make use of a BiLSTM which extends the LSTM architecture \cite{hochreiter1997long} by processing the sequence in both forward and backward temporal directions.

\begin{figure*}[bt!]
         \centering
         \includegraphics[width=1\textwidth]{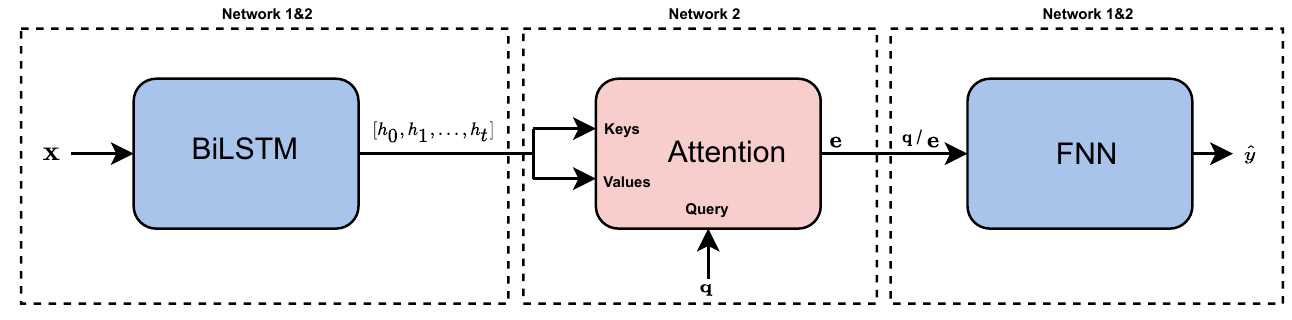}
         \vspace*{-5mm}
         \caption{Structure of the basic BiLSTM (Network 1) and its attention variant (Network 2), with shared components indicated.}
         \label{fig:network}
\end{figure*}

A high-level diagram of the network is shown in Figure \ref{fig:network}. 
The single BiLSTM layer has a 32-dimensional hidden state whereby the final outputs in both directions
are concatenated to form $\mathbf{q}$. This embedding is then passed through a small feed-forward network with a 32-dimensional hidden layer and ReLU activations followed by an output layer. We include dropout before the first linear layer with a probability of $0.5$. Lastly, to account for the unbalanced nature of our dataset, we use weighted cross-entropy as our loss function.

\subsection{BiLSTM-Att}
\vspace*{-1mm}

The development of the attention mechanism \cite {vaswani2017attention} has revolutionised deep learning research. With a focus on acoustic classification, attention-based architectures achieve near state-of-the-art results on tasks such as the Google speech commands dataset \cite{rybakov2020streaming, de2018neural}. In addition, the intuitive nature of the architecture allows for analysis of what the network is learning, reducing the black-box notion commonly associated with deep learning.

We develop an attention-based model by integrating an attention layer into the above BiLSTM architecture. Instead of passing $\mathbf{q}$ directly to the fully connected network as is done in the basic BiLSTM architecture, the attention mechanism uses $\mathbf{q}$ as the query and outputs a weighted average (by the attention score) of all the BiLSTM outputs, thereby allowing the single output vector to capture information from the temporal regions most relevant for classification and suppress information from unimportant regions in time.

We design this architecture bearing in mind the fact that it will be used to aid in the understanding of the acoustic signature of a TB cough. 
Accordingly, a new loss term is introduced that encourages the embedding layer of the network (the output of the attention block) to generalise across patients of the same TB status. 
This is performed to inhibit our subsequent model analysis to be confounded by attributes learned irrelevant to TB cough classification, namely patient identity, an attribute present in cough \cite{zhang2017speaker}. 
This is accomplished by incorporating a GE2E loss term which was originally proposed to determine speaker embeddings by encouraging the network to keep embeddings close together when from the same target speaker, and further apart for different speakers~\cite{wan2018generalized}.

We consider TB and $\overline{\text{TB}}$ coughs to represent two respective ``speakers". 
Hence the similarity between the embedding centroids of different patients with the same TB status is maximised, whilst minimising the similarity between embeddings of the TB and $\overline{\text{TB}}$ classes. 
The combined loss function used to train our network is given in Equation \ref{eq:lge2e}, where $B$ is the batch size.

\vspace*{-1mm}
\begin{equation}
\mathcal{L} = - \frac{1}{B}\left( \sum^{B}_{b} \beta \cdot \mathbf{y_{b}}\cdot \log({\mathbf{\hat{y}_{b}}}) + \alpha \sum_{j, i} L_{GE2E}(\mathbf{e}_{ij})\right)
 \label{eq:lge2e}
\end{equation}

Here the first term is standard weighted cross-entropy where $\mathbf{y_{b}}$ and $\mathbf{\hat{y}_{b}}$ are the vectors of ground truth and predicted probabilities respectively (where the dimension is the number of classes i.e. two) for a given cough in the batch and $\mathbf{\beta}$ is a vector of class weights (constant throughout training). The weight for the under-sampled class is set to $1$, while for the over-sampled class it is the ratio of its occurrence in the training set to the total number of samples. In the second term, $\alpha$ is a regularisation parameter, $L_{GE2E}$ is the function that computes the GE2E loss for a specific embedding in a given batch, and $\mathbf{e}_{ij}$ is the embedding vector of the $i^{th}$ cough from the $j^{th}$ class.

\section{Experimental procedure}

With the exception of LR, which is trained using the standard \texttt{scikit-learn} recipe \cite{scikit-learn}, models are trained for 15 epochs with a learning rate of $\num{1e-4}$ and batch size of $128$. 
After training all 4 folds, the mean development AUC is computed for each epoch. 
The models from the epoch with the highest mean development AUC are selected, and at test time are ensembled. 
The decision threshold used for classification was $\bar{\gamma}$, the mean of the decision thresholds $\gamma_n$ that result in the EER for each fold. 
Hence the decision threshold was chosen on the basis of the EER as in previous work \cite{botha2018detection, pahar2021automatic}. 
We note that it might be possible to improve performance if a strategy that chooses this threshold to optimise, for example, sensitivity and specificity, is adopted. 
However, we leave this investigation for future work.

\subsection{Data Augmentation and feature extraction}
\vspace*{-1mm}

We experimented with 3 data augmentation techniques: SpecAugment \cite{park2019specaugment}, random insertions and deletions, and speed-perturbation \cite{ko2015audio}. Initial experiments indicated that only speed perturbation was effective, and hence report only this form of augmentation. We use warping factors of $\{0.9, 1.0, 1.1\}$ resulting in a 3-fold increase in the size of the dataset.
We considered three types of acoustic feature: mel-spectrograms, linear filter-bank energies, and MFCCs (with appended velocity and acceleration, as well as cepstral mean and variance normalisation). The former was found to perform best for all architectures, with the ideal number of filter banks being 180, 128, and 80 for LR, the BiLSTM and BiLSTM-Att model respectively.
Analysis was based on 2048-sample frames, with successive frames overlapping by 1536 samples (i.e. a frame-skip of 512). Variations in frame length and frame-skip were not considered in this work. All recordings were down-sampled to $44.1$kHz before feature extraction.

\subsection{Idealised coughs through neural style transfer} 
\label{sec:trans}
\vspace*{-1mm}

In an attempt to understand what the network is learning in order to distinguish between TB and $\overline{\text{TB}}$ coughs, we employ a technique similar to that used in neural style transfer \cite{gatys2015neural} to synthesise an \textit{idealised} cough for each class. This is accomplished by first defining a 2D parameter matrix $\hat{\mathbf{x}}\in\mathbb{R}^{n \times d}$ (initialised to zeros) that will represent the input to the network, where $d$ is the size of the acoustic feature vector seen by the network during training, and $n$ is the number of frames. In this case we select $n=80$. Next, we perform training as before, but instead of fixing the network input and output and optimizing the weights, we fix the weights and train $\hat{\mathbf{x}}$. This allows the discovery of the input $\hat{\mathbf{x}}$ that best leads to the output class $y$ for the trained weights, i.e. an idealised input cough feature representation for the output class in question.

\begin{figure}[h!]
    \centering
     \begin{subfigure}[bt]{0.45\textwidth}
         \centering
         \includegraphics[width=\textwidth]{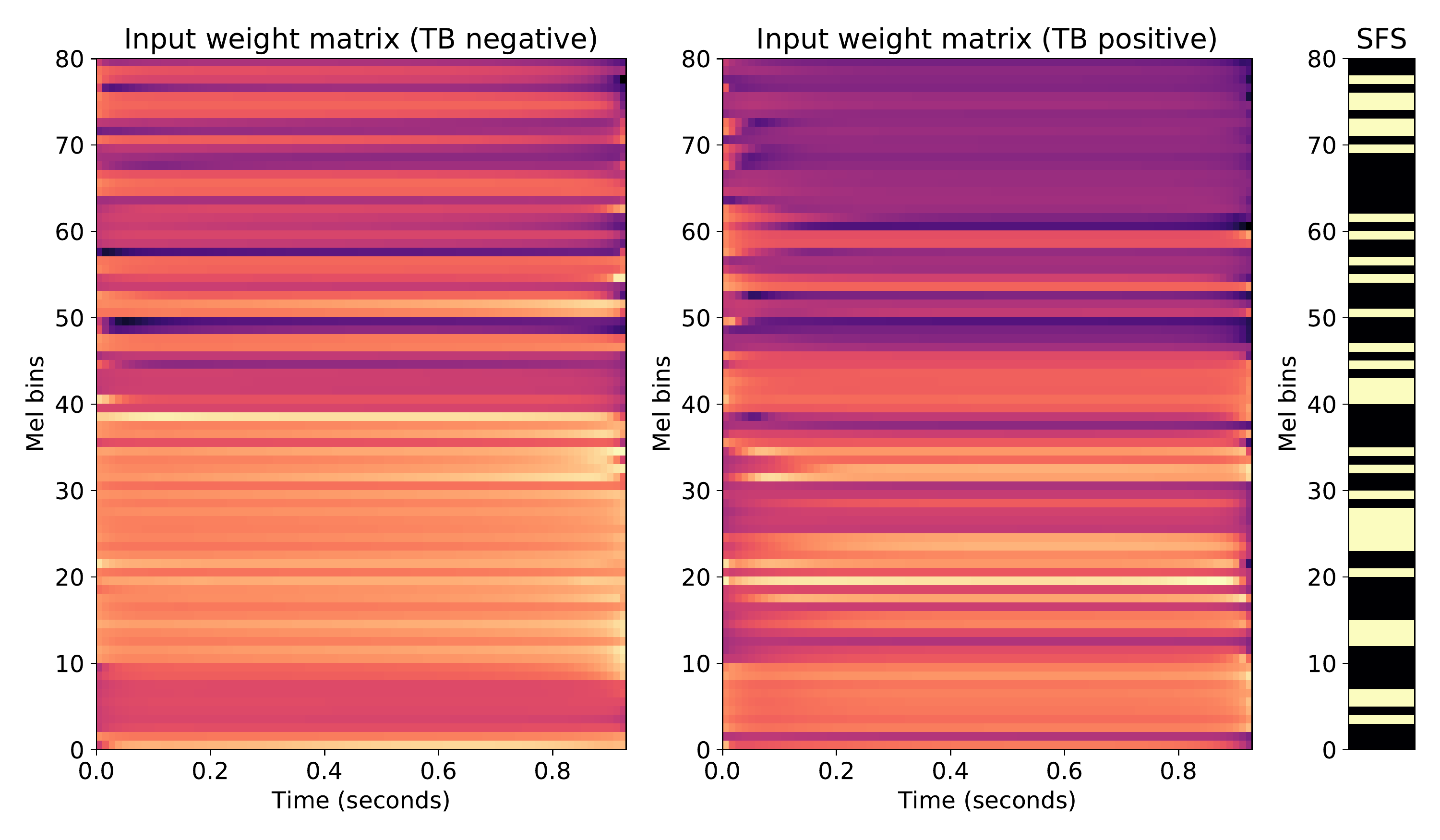}
         \vspace*{-5mm}
         \caption{Idealised mel-spectrograms for TB negative (left) and positive (center) coughs, with bins identified by SFS   shown in yellow (right).}
         \label{subfig:ideal_coughs}
     \end{subfigure}
     \begin{subfigure}[bt]{0.45\textwidth}
         \centering
         \includegraphics[width=\textwidth]{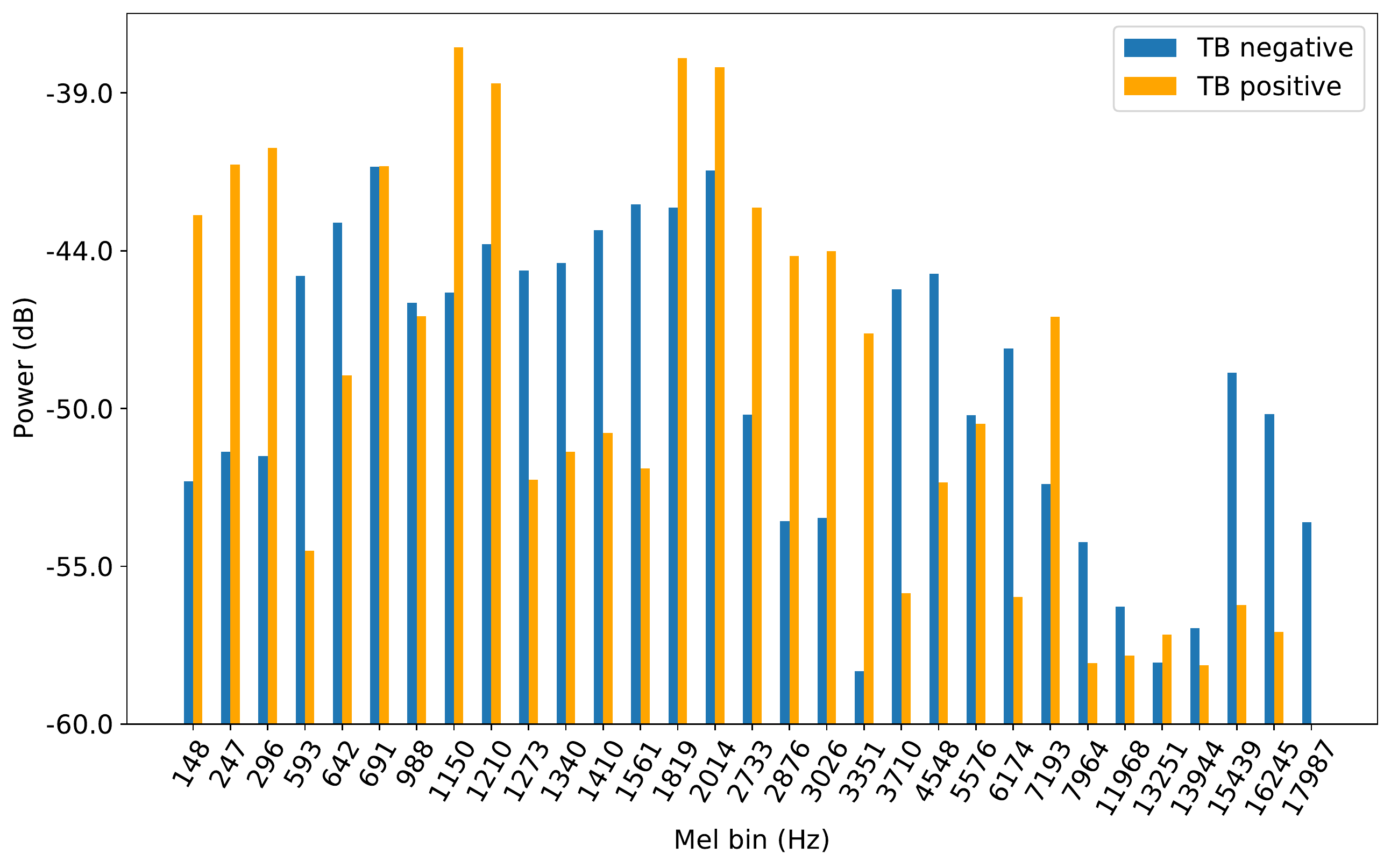}
         \vspace*{-5mm}
         \caption{Mean power of each idealised mel-spectrogram for the frequency bins deemed most important by SFS.}
         \label{subfig:mean_powers}
     \end{subfigure}
     \caption{Idealised mel-spectrograms and mean spectral power.}
\end{figure}
\vspace*{-2mm}

\section{Experimental results and discussion}
We present classification performance for the various classifiers investigated and discuss our findings. In addition, we analyse the idealised cough mel-spectrograms produced when applying our adaptation of neural style transfer and observe the attention weights to infer the spectral and temporal regions that the classifier finds most useful for classification.

\subsection{Classification}
\vspace*{-1mm}

\begin{figure*}[h!]
         \centering
         \includegraphics[trim={0 0 0 16cm},clip,width=1\textwidth]{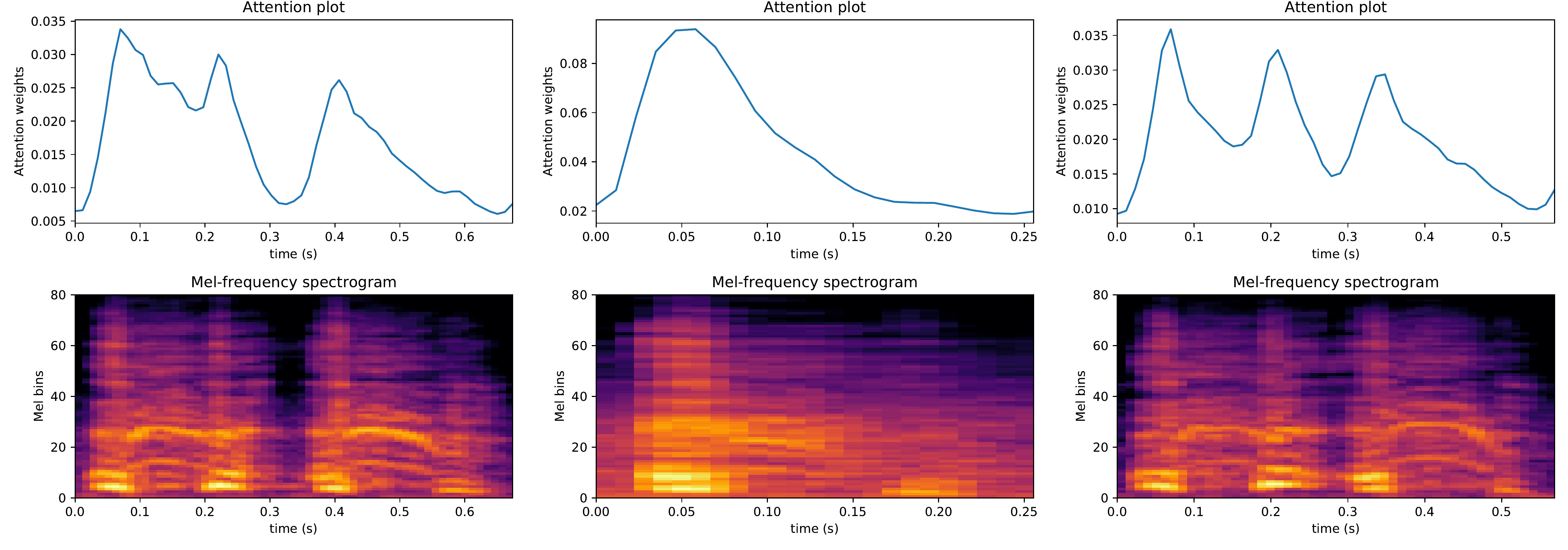}
         \caption{Various cough mel-spectrograms, all of which originate from the same patient, and their respective attention weights shown above them.}
         \label{fig:attention_plots}
\end{figure*}

We present development set performance for each considered architecture in Table \ref{tab:dev} and the associated test set performance in Table \ref{tab:test}. A substantial increase in classification performance over the LR baseline is observed with regards to the basic BiLSTM model with all metrics either matched or improved upon, most notably the specificity. 
The test set AUC for both the basic BiLSTM and its attention variant are comparable, but there is a large discrepancy in the remaining metrics. This indicates that the decision threshold was not optimal and more robust alternatives to using the EER should be explored in future work.

When inspecting the effect of applying SFS on the two deep architectures, interesting observations can be made. We note a substantial reduction in the standard deviation of the EER-based thresholds determined for the BiLSTM-Att architecture ($0.175$ before and $0.070$ after SFS) whilst the opposite is observed for the BiLSTM without attention. Despite achieving the highest test AUC, an increase in decision threshold standard deviation was observed (from $0.108$ to $0.155$). An increase in the variability of the decision threshold between folds indicates poorer generalisation. 
Conversely, with the BiLSTM-Att architecture, better generalisation across the folds is observed which is evident in the reduced standard deviation of the decision threshold. This is especially important with the implementation of a TB screening tool in mind, where model generalisation will be key.

\vspace{-2mm}
\begin{table}[tb]
        \centering
        \caption{Mean and standard deviation of the area under the ROC curve (AUC) and the EER decision thresholds ($\bar{\gamma}$) observed during 4-fold cross validation.}
         \begin{tabular}{l c c}
          \toprule
          Model & $\bar{\gamma}$ & AUC\\
          \midrule
           LR (\textit{baseline}) \cite{botha2018detection, pahar2021automatic}  & $0.272\pm 0.103$ & $0.701 \pm 0.127$\\
           BiLSTM  & $ 0.534 \pm 0.108$ & $0.777 \pm 0.094$\\
           BiLSTM  (SFS) & $0.603 \pm 0.155$ & $0.919 \pm 0.081$\\
           BiLSTM-Att & $0.460 \pm 0.175$ & $0.873 \pm 0.054$\\
           BiLSTM-Att (SFS) & $0.568 \pm 0.070$ & $0.900 \pm 0.092$\\
          \bottomrule
        \end{tabular}
     \label{tab:dev}
\end{table}
\begin{table}[tb]
        \centering
        \caption{Test set performance for the models listed in Table \ref{tab:dev}, evaluated through various metrics: sensitivity, specificity, accuracy and area under the curve.}
         \begin{tabular}{l c c c c}
          \toprule
          Model & Sens & Spec & Acc & AUC\\
          \midrule
           LR (\textit{baseline}) \cite{botha2018detection, pahar2021automatic} & \textbf{0.889} & 0.625 & 0.720 & 0.769 \\
           BiLSTM & \textbf{0.889} & 0.750 & \textbf{0.800} & 0.821 \\
           BiLSTM (SFS) & 0.667 & 0.750 & 0.720 & \textbf{0.862} \\
           BiLSTM-Att & 0.778 & 0.625 & 0.680 & 0.822 \\
           BiLSTM-Att (SFS) & 0.778 & \textbf{0.813} & \textbf{0.800} & 0.850 \\
          \bottomrule
        \end{tabular}
     \label{tab:test}
\end{table}
\vspace{-1mm}
\subsection{Analysis and interpretation}
\vspace*{-1mm}

Figure \ref{subfig:ideal_coughs} depicts the idealised coughs synthesised by the BiLSTM-Att network using the neural style transfer method described in Section \ref{sec:trans}. Clear differences between idealised $\overline{\text{TB}}$ and TB cough mel-spectrograms are observed. This is further illustrated by comparing the mean power of these idealised coughs as a function of the frequencies determined to be most important by SFS, as shown in Figure \ref{subfig:mean_powers}. For the idealised TB cough, we observe generally higher power at lower frequencies ($<500\si{Hz}$) and the mid-band range of $1.8\si{kHz} - 3.3\si{kHz}$ whereas the $\overline{\text{TB}}$ cough has higher power between $1.2\si{kHz} - 1.8\si{kHz}$ and frequencies greater than $3.7\si{kHz}$, which include frequencies far outside the typical range of human speech \;\; ($>8\si{kHz}$). 
In Figure \ref{fig:attention_plots} we plot the attention weights as a function of time for three cough mel-spectrograms. We observe large importance being placed on regions where the signal has a high power and a large bandwidth, which coincide with the initial bursts of energy for each coughing episode. Whilst only three examples are shown, these observations were made in general. This high energy portion of the coughing sound originates from the lung itself, in particular, the bronchi \cite{korpavs1996analysis}. It therefore appears that, whilst TB can manifest in all regions of the respiratory tract, the model is relying on some change in the sound produced inside the lungs of TB and $\overline{\text{TB}}$ patients. Further research is necessary to deduce what the physiological causes of this difference in the audio signals could be.

\vspace{-2mm}
\section{Conclusion}
\vspace{-1mm}
 In this work, we explored the use of recurrent networks for TB cough classification and use these trained networks to identify and interpret important cough characteristics in both frequency and time. A BiLSTM architecture is shown to improve on previous research, achieving a sensitivity and specificity of 0.89 and 0.75 respectively. This indicates that deeper architectures are viable for TB cough classification, and can improve upon previous state-of-the-art for TB screening. Furthermore, we show that by incorporating frugal feature selection our proposed attention-based architecture exhibits substantially better generalisation across folds than the other considered architectures. This is an important observation for future work, in which datasets will include many more recording domains and associated variability.
  Utilizing an attention architecture, the importance of certain temporal regions in the cough signal could be visualised. It was observed that the initial voiced regions of cough were the most important for classification. Moreover, by employing a neural style transfer technique, idealised TB negative and positive coughs were synthesised. Subsequent inspection revealed stark differences between energy content in specific frequency bands. In addition to providing new insights into the aspects of a tuberculosis cough that are important for classification, this provides evidence that the TB signal being learnt does indeed originate in the lungs. In future work, we look forward to evaluating our architectures on larger datasets currently being collected \cite{cagetb2022}.
\vspace{-3mm}
\section{Acknowledgements}
The authors gratefully acknowledge funding from the EDCTP2 programme supported by the European Union (grant RIA2020I-3305, CAGE-TB).

\bibliographystyle{IEEEtran}

\bibliography{ms}

\end{document}